\documentclass[12pt]{article}
\usepackage{cite}
\usepackage[pdftex]{graphicx}
\usepackage{moreverb}
\usepackage{amssymb}                    
\usepackage{amscd}                   
\usepackage{wasysym} 
\usepackage{makeidx} 
\def\FF#1{\begin{eqnarray} #1 \end{eqnarray}}
\newcommand{\ML}{\textsc{Matlab}\ }
\newcommand{\re}{{\rm e}}
\newcommand{\ri}{{\rm i}}
\newcommand{\rd}{{\rm d}}
\newcommand{\nn}{\nonumber}
\newcommand{\ah}{\ensuremath{\hat a}}
\newcommand{\ad}{\ensuremath{\hat a^\dagger}}

\newcommand{\nh}{\ensuremath{\hat n}}
\newcommand{\ro}{\ensuremath{\hat \rho}}
\newcommand{\xh}{\ensuremath{\hat x}}
\newcommand{\ph}{\ensuremath{\hat p}}
\newcommand{\Ah}{\ensuremath{\hat A}}

\newcommand{\Ih}{\ensuremath{\hat I}}

\newcommand{\xb}{\ensuremath{\beta}}
\newcommand{\xa}{\ensuremath{\alpha}}
\newcommand{\yy}{\ensuremath{\sigma}}
\newcommand{\cc}{\ensuremath{c}}
\newcommand{\uu}{\ensuremath{u}}
\newcommand{\bb}{\ensuremath{b}}
\newcommand{\vv}{\ensuremath{v}}
\newcommand{\zz}{\ensuremath{z}}
\newcommand{\dd}{\ensuremath{\delta}}
\newcommand{\tr}{\ensuremath{{\rm \,trace\,}}}
\newcommand{\omti}{\ensuremath{\tilde \omega}}
\newcommand{\fti}{\ensuremath{f}}
\newcommand{\Fti}{\ensuremath{\tilde f}}

\newcommand{\cF}{\ensuremath{{\cal F}}}

\begin{document}
\title{Lindblad dynamics of the damped and  forced quantum harmonic oscillator}
\author{H. J. Korsch\thanks{Electronic address: h.j.korsch@gmail.com}\\
FB Physik, Technische Universit\"at Kaiserslautern\\ D-67653 Kaiserslautern, Germany}
%
\maketitle
\begin{abstract}
The quantum dynamics of a damped and forced harmonic oscillator  is 
investigated in terms of a Lindblad master equation. Elementary algebraic 
techniques are employed allowing for example to analyze the long time behavior, i.e.~the quantum limit cycle. The time evolution of 
various expectation values is obtained in closed form as well as
the entropy and the Husimi phase space distribution. We also discuss 
the related description in terms of a non-Hermitian Hamiltonian.
\end{abstract}
\section{Introduction}
\label{s-int}
In classical mechanics, the damped and driven harmonic oscillator is one of the 
paradigmatic systems
discussed in elementary physics lectures, modeling for example a mass
attached to a linear spring in the gravitational field. The dynamics
is linear and allows a closed form solution, which then serves as a basic
model for analyzing more general and, of course, much more demanding
general dynamical systems.
The harmonic oscillator also constitutes a basic building block in elementary
quantum mechanics, however almost exclusively in context
with Hermitian dynamics, which is also in most cases restricted to systems with
no explicit time-dependence, though also a time-depended force can be
treated quite elementary, as for instance in the book by Louisell 
\cite[sect.~3.11]{Loui73}. The damped quantum harmonic oscillator is, however,
addressed very rarely in elementary texts. Exceptions are, e.g., the recent
paper by Fujii \cite{Fuji12} describing in much detail the dissipative quantum dynamics 
in terms of the Lindblad master equations, however for the force-free case.
This system is also described in the article \cite{Lamp16} as well as in 
lecture notes by  Englert and Morigi \cite{Eng02}, where
also a quantum-optical  system with periodic driving is studied, 
describing a  decaying photon field excited by periodic kicking.

An elementary study of the Lindblad dynamics of a forced harmonic oscillator seems
to be missing, however, and here we will try to fill this gap. We 
follow the analysis in \cite{Fuji12}, describing briefly these results 
for the non-driven system in section \ref{s-lindblad}. In this case the general 
solution can be constructed and the density operator is expressed in
closed form. This is, however, not fully achieved in the present analysis 
of the driven oscillator in section \ref{s-special}. We will derive a special class of solutions using similar techniques as in our previous study of a related model \cite{92dqs}. 
Nevertheless we will be  able to describe some important features of the 
dynamics, as for example the density operator for the quantum limit cycle.
 Section \ref{s-nh} is devoted to
a related description of quantum damping in terms of non-Hermitian Hamiltonians,
in the present case a harmonic oscillator with a complex frequency. For convenience
of the reader we begin with a brief repetition of the well-known classical dynamics.
We will use units with $\hbar=1$ throughout.
\section{The classical oscillator}
\label{s-classical}
Let us briefly recall the damped and driven harmonic oscillator  
in classical mechanics. The equation of motion
\FF{\ddot x +2\gamma\dot x +\omega_0^2 x=\Fti(t)\label{dgl-cl}}
models a harmonic oscillator with frequency $\omega_0$ under a time-dependent
force $\Fti(t)$, which will be assumed to be harmonic,
\FF{\Fti(t)=\Fti_0\cos \Omega t\,,\label{force}}
 in the following. The interaction with an environment is modeled by the term $2\gamma\dot x$
where the constant $\gamma>0$ describes the friction. 
For a free motion with vanishing force the general solution is given by
\FF{x(t)=x_+\re^{+\ri \omega t-\gamma t}+x_-\re^{-\ri \omega t-\gamma t}\,,
\label{free-sol-cl}}
where the constants $x_+$ and $x_-$ are determined by the initial conditions.
The frequency 
\FF{\omega=\sqrt{\omega_0^2-\gamma^2}\label{omega-cl}}
is real valued in the underdamped case, $\gamma<\omega_0$, resulting
in an oscillatory motion, whereas $\omega$ is imaginary for strong damping,
$\gamma>\omega_0$,
and both terms in (\ref{free-sol-cl}) are exponentially
decaying. In any case, the free solution finally approaches $x=0$.
For a harmonic force (\ref{force}) one can construct a special solution,
the limit cycle
\FF{x^{\rm (\ell c)}(t)=A\cos(\Omega t +\phi)\,,\label{x0-cl}}
where the motion follows the external force, shifted by a phase 
\FF{\phi=-\arctan\frac{2\gamma\Omega}{\omega_0^2-\Omega^2}}
with an amplitude
\FF{A=\frac{\Fti_0}{\sqrt{(\omega_0^2-\Omega^2)^2+(2\gamma\Omega)^2}}.}
The general solution is given
by a sum of (\ref{free-sol-cl}) and (\ref{x0-cl}). Then for
all initial conditions the solution $x(t)$ converges to the limit cycle (\ref{x0-cl}).
As a function of the external frequency $\Omega$, the
limit cycle shows the celebrated resonance behavior, where the amplitude $A$
shows a maximum of height $A_R=\Fti_0/(2\gamma\sqrt{\omega_0^2-\gamma^2})$
at the resonance frequency $\Omega_R=\sqrt{\omega_0^2-2\gamma^2}$ for
 $\omega_0^2>\gamma^2$ and the phase shift $\phi$ changes from
 $\phi=0$ for $\Omega=0$ to $\phi=-\pi$ for large $\Omega$.
If we identify $\dot x$ with the momentum $p$
(assuming unit mass) the limit cycle appears in phase space as an ellipse, namely
\FF{p^2/\Omega^2 +x^2=A^2\,.\label{limitcycle-cl}}
\section{The Lindblad master equation}
\label{s-lindblad}
The quantum dynamics of our system is described by the
Lindblad master equation
\FF{&&\frac{\rd\ro}{\rd t}={\mathcal L} \ro = -\ri\, \big[\hat H,\ro\,\big]\nn\\
   &&\qquad +
   \frac{\mu}{2}\,\big(2 \ah \ro\ad -\ad \ah\ro- \ro\ad\ah\big)
   +\frac{\nu}{2}\,\big(2 \ad \ro\ah -\ah \ad\ro- \ro\ah\ad\big),
\label{Lindblad}}
where the
 $\mu$- and $\nu$-terms represent (homogeneous) loss and
 (inhomogeneous) pumping of particles, respectively. Note that often these coefficients
 are written as $\mu=\gamma_0(1+\overline n)$ and 
 $\nu=\gamma_0\,\overline n$, where $\overline n=\nu/(\mu-\nu)$ is the thermal population of the environment. In the following it will be convenient
to rewrite the damping coefficients as 
\FF{\mu=\gamma'+\gamma\  ,\quad \nu=\gamma'-\gamma\,,} 
where $\gamma'$ and $\gamma$ are denoted as diffusion and dissipation
constants \cite{Piil06}. 
We will assume $\mu>\nu \ge 0$ so that $\gamma'=(\mu+\nu)/2$ and
$\gamma=(\mu-\nu)/2$ are  positive.
A motivation and some
applications of this evolution equation can be found in \cite{Eng02}
or \cite{Breu02}, where it is shown that this evolution equation can, for
example, describe the damping of an electromagnetic field mode
inside a cavity. The modes outside the cavity in a thermal reservoir cause a damping 
with rate $\gamma$.

It may be recalled that the density operator $\ro(t)$ is a positive operator with unit trace and the Lindblad evolution conserves the trace and the positivity of $\ro$. 
The expectation value of an
observable $\Ah$ is given by 
\FF{\langle \Ah\rangle_t=\tr (\Ah \ro(t))\,,}
where we have assumed that $\Ah$ is not explicitly time-dependent.
If $\hat H$ is Hermitian, we have 
\FF{\langle \Ah^\dagger\rangle_t=\langle \Ah\rangle^*_t\,.
\label{Adagger-av}}

The  Hamiltonian of the forced harmonic oscillator is 
 \FF{\hat H=\omega\big(\ad\ah+1/2\big)- \fti^*(t)\,\ad-\fti(t)\,\ah\label{Hamiltonian}}
with time-independent real $\omega$,
where $\ad$ and $\ah$ are the familiar bosonic creation and
destruction operators with commutator $|\ah,\ad]=1$. We will also
use the number operator $\nh=\hat a^\dagger\hat a$, whose eigenvalues $n=0,\,1,\,\ldots$ are the number of oscillator quanta, sometimes also denoted as the number of
photons or simply particles. This Hamiltonian models, for example, the coherent excitation of the cavity field by a monochromatic laser. 

The Hamiltonian is Hermitian and can be rewritten as
\FF{\hat H=\omega\big(\ad\ah+1/2\big)- \Fti(t)\,\xh\label{Hamiltonian2}}
 for a real valued force, using
\FF{\xh=(\ad+\ah)/\sqrt{2\omega}, \quad \ph=\ri(\ad-\ah)\sqrt{\omega/2} \quad 
\textrm{and}\quad \Fti(t)=\sqrt{2\omega}\,\fti(t)\,.}
In the following we will consider some special cases in more detail,
namely 
a force-free system, $\fti(t)=0$, which is time-independent and 
allows  the construction of a general analytical solution $\ro(t)$, and
a harmonic driving $\fti(t)=\fti_0\,\cos \Omega t$ with real valued
amplitude $\fti_0$. Here a closed form solution can be derived for 
a special set of initial distributions. 

 Without the forcing term, the  
density $\ro(t)$ approaches for $\nu>0$ in the long time limit the steady state distribution \cite{Eng02}
\FF{\ro^{\rm (ss)}=(1-\nu/\mu)\,\re^{\,\ln (\nu/\mu)\,\nh}\label{ro-stst}}
and the mean value $\langle \nh\rangle$ is equal to
$\overline{n}=\nu/(\mu-\nu)=\nu/2\gamma$, which agrees
with the mean number of photons in the environment
if the system models a photon field in a cavity. For $\nu=0$,
where we have no gain,  the steady state distribution
is given by the oscillator ground state $\ro^{\rm (ss)}=|0\rangle\langle 0|$.
In the following we will be interested
in the Lindblad dynamics of a forced harmonic oscillator, where the force
is time-dependent with emphasis on a harmonic driving.

The Lindblad equation (\ref{Lindblad}) can, of course, be solved numerically
(a simple \ML program can be found in \cite{16easy2}), but the dynamics shows 
various features, which are not directly obvious and a closed form exact
solution is in many cases preferable, in particular for such a quite
elementary system. In the following we will discuss this in some detail.
A general solution describing the density operator $\ro(t)$ for any given $\ro_0$
 has been presented recently by Fujii \cite{Fuji12}
for the non-driven case and some of these results will be given below. 
Solutions, however, can also be found for
a driven system, as for instance for the expectation values of
position, momentum and energy, or for a special class of density
operators as shown in the following sections. 
\subsection{The force-free oscillator}
\label{s-sol-free}
For a free evolution, $\fti(t)=0$, the Hamiltonian is time-independent,
which considerably simplifies the analysis. A full solution has been
presented recently by Fujii \cite{Fuji12} based on previous work 
\cite{Endo08,Fuji08}. Here the dynamics of the density operator is mapped
on a vector evolution using the Kronecker product (see also the book by
Steeb and Hardy \cite{Stee11} as well as \cite{Amsh15}).\footnote{Note that 
such a technique 
has also been developed as a Liouville space formulation by Ban \cite{Ban92}.}
This derivation is heavily based on techniques
from operator algebra and here we only state the final result obtained 
in \cite{Fuji12}, namely
\FF{\ro(t)&=&\frac{\re^{\gamma t}}{F(t)}\sum_{j=0}^\infty
\frac{G(t)^j}{j!} \ah^{\dagger\,j}\,\re^{\{-\ri\omega t -\ln F(t)\}\nh}
\times\nn\\
&&\qquad \qquad \Big\{\sum_{k=0}^\infty \frac{E(t)^k}
{k!}\ah^k\ro_0\,\ah^{\dagger k}\Big\}
\,\re^{\{+\ri\omega t -\ln F(t)\}\nh}\,\ah^j\,,\label{sol-fujii}}
where the three functions $E(t)$, $F(t)$ and $G(t)$ are solutions of some
 first order differential equations with $F(0)=1$  and $E(0)=G(0)=0$. 
Let us note, for example, that $G(t)$ solves 
 the Riccati equation
 \FF{\dot \uu+(\mu+\nu)\uu-\mu\uu^2=\nu\,,\label{riccati}}
for an initial condition $\uu(0)=0$. As shown in \cite{Fuji12} these 
functions are given by
\FF{F(t)&=&\cosh \gamma t+(\gamma'/\gamma)\sinh \gamma t \\[1mm]
E(t)&=&\frac{\mu}{\gamma F(t)}\sinh \gamma t\ ,\quad
G(t)=\frac{\nu}{\gamma F(t)}\sinh \gamma t\,.\label{EtGt}}
\vspace*{2mm}

The general solution (\ref{sol-fujii}) simplifies for special cases (see \cite{Fuji12} for details):\\[1mm]
$(\bullet)$ For $\nu=0$  one obtains
 \FF{\ro(t)=\re^{-(\mu/2+\ri\omega)t\,\nh}
 \Big\{\sum_{k=0}^\infty \frac{E(t)^k}
{k!}\ah^k\ro_0\,\ah^{\dagger k}\Big\}\re^{-(\mu/2-\ri\omega)t\,\nh}\,.
\label{sol-fujii0}}

\noindent
$(\bullet)$ If $\nu\ne 0$ and the
initial state  is the oscillator ground state $\ro_0=|0\rangle\langle 0|$, we have
\FF{\ro(t)=(1-G(t))\,\re^{\,\ln G(t)\,\nh}\label{ro-G-n0}}
and the mean value of $\nh$ is given by
\FF{\langle \nh\rangle_t=\tr (\nh \ro(t))
=\frac{\nu}{2\gamma}\Big(1-\re^{-2\gamma t}\Big)\,.\label{nav-n-0f-0}}
Here the system is for  all times in a thermal state, i.e.~a  canonical
distribution. In the long time limit we clearly see with $G(t)\rightarrow \nu/\mu$
 the convergence toward the steady state distribution
(\ref{ro-stst}). 
\vspace*{2mm}

\noindent
$(\bullet)$ For $\nu \ne 0$ and a coherent initial state $\ro_0=|\xa_0\rangle\langle \xa_0|$
the solution is given by 
\FF{\ro(t)=\re^{c(t)-\log G(t)\big(\xa(t)\ad+\xa^*(t)\ah -\nh\big)}
\label{rho-coh-0}}
with
\FF{c(t)=|\xa(t)|^2\log G(t) +\log(1-G(t))
\ ,\quad \xa(t)=\xa_0\re^{-(\gamma+\ri\omega)t}\,.\label{rho-coh-beta}}
(compare eq.~(4.3) in \cite{Fuji12}), or, written in product form using the disentangling relation given in eq.~(\ref{koeff-transf}) below, 
\FF{\ro(t)=\re^{\tilde c(t)}\re^{(1-G(t))\xa(t)\ad}
\re^{\log G(t)\nh}\re^{(1-G(t))\xa^*(t)\ah}
\label{rho-coh-0x}}
with
\FF{\tilde c(t)=\log (1-G(t)) -|\xa(t)|^2(1-G(t))\,.}
\vspace*{0mm}

\noindent
$(\bullet)$ Finally, for $\ro_0=|\xa_0\rangle\langle \xa_0|$
and $\nu=0$, we find from (\ref{sol-fujii0})
\FF{\ro(t)&=&\re^{(1-\re^{-\mu t})|\xa_0|^2}\,\re^{-(\mu/2+\ri\omega)t\,\nh}
\,|\xa_0\rangle\langle \xa_0|\,\re^{-(\mu/2-\ri\omega)t\,\nh}\nn\\
&=&|\xa(t)\rangle\langle\xa(t)|
\quad \textrm{with}\quad \xa(t)=\xa_0\,\re^{-(\mu/2+\ri\omega)t}
\label{rho-coh-nu0}}
(compare eq.~(\ref{rho-coh-nu0-A}) in appendix \ref{ss-coherent}),
i.e.~the system remains in a coherent state for all times and the mean values 
of  $\ah$ and $\nh$ decay as
\FF{\langle \ah\rangle_t=\xa(t) \ ,\quad
\langle \nh\rangle_t=|\xa(t)|^2=|\xa_0\,\re^{-(\mu/2+\ri\omega)t}|^2
=n_0\,\re^{-\mu t}\label{nav-coh-0}}
with $n_0=|\xa_0|^2$.
\subsection{Mean value evolution}
\label{s-mean}
Returning to the forced oscillation, we will first consider the time evolution 
of some expectation values. For the operators $\ah$, $\ad$ and $\nh$ this can be achieved quite
easily. Let us demonstrate this in some detail for $\ah$. The time derivative
of $\langle \ah\rangle_t=\tr (\ah\ro(t))$ yields 
\FF{&&\frac{\rd \langle \ah\rangle}{\rd t}=\tr \big(\ah\, \frac{\rd \ro}{\rd t}\Big) \nn\\
&&=-\ri \omega\tr \big(\ah\big[\ad\ah,\ro\big]\big) 
+\ri \tr  \big(\ah\,\big[\fti^*(t)\ad+\fti(t)\ah,\ro\big]\big)\\[2mm] 
&&+\frac{\mu}{2} \tr \big(2\ah\ah\ro\ad-\ah\ad\ah\ro-\ah\ro\ad\ah\big)
+\frac{\nu}{2} \tr \big(2\ah\ad\ro\ah-\ah\ah\ad\ro-\ah\ro\ah\ad\big).\nn
}
This can be simplified using $\tr (\Ah\hat B)=\tr (\hat B\Ah)$  with the result
\FF{\frac{\rd \langle \ah\rangle}{\rd t}=-(\ri\omega+\gamma)\langle \ah\rangle
+\ri \fti^*(t)\,,\label{a-av-dgl}}
which can be integrated as
\FF{\langle \ah\rangle_t=\re^{-(\ri \omega +\gamma)t}\langle \ah\rangle_0
+\cF(t)\label{a-sol-gen}}
with
\FF{\cF(t)=\ri\!\int_0^t\rd t' \fti^*(t')\,\re^{-(\ri \omega +\gamma)(t-t')}\,.
\label{cFt}}

For a time-periodic force $f(t)=f(t+\Omega/2\pi)$ one can evaluate the integral
(\ref{cFt}) by Fourier expansion
\FF{f(t)=\sum_k c_k\,\re^{i k \Omega t}\label{fou-ex}}
with the result
\FF{\cF(t)=\sum_k c_k^*\,
\frac{\re^{\ri k\Omega t}-\re^{-(\ri\omega+\gamma)t}}{\omega+k\Omega-\ri\gamma}\,.
\label{fou-F}}
In the same way one can derive an equation for $\langle \ad\rangle_t$, which
agrees with ${\langle \ah\rangle}^*_t$ (compare eq.~(\ref{Adagger-av})).

For vanishing driving we have, of course, 
\FF{\langle \ah \rangle_t=\re^{-(\ri \omega +\gamma)t}\langle \ah\rangle_0}
and for  a harmonic driving $\fti(t)=\fti_0\cos\Omega t$ and one obtains
\FF{\cF(t)=\frac{\fti_0}{2}\,\Big(
\frac{\re^{\ri \Omega t}-\re^{-(\ri \omega +\gamma)t}}{\omega+\Omega-\ri\gamma}
+\frac{\re^{-\ri\Omega t}-\re^{-(\ri \omega +\gamma)t}}{\omega-\Omega-\ri\gamma}\Big)\,,
\label{at-sol}}
which then provides the solution for the expectation values of
position $\xh=(\ad+\ah)/\sqrt{2\omega}$ and  momentum $\ph=\ri\sqrt{\omega/2\,}\,(\ad-\ah)$.
In the limit $t\rightarrow\infty$ we have
\FF{\cF(t) \rightarrow\cF^{\rm (\ell c)}(t)=
\frac{\fti_0}{2}\,\Big(
\frac{\re^{\ri \Omega t}}{\omega+\Omega-\ri\gamma}
+\frac{\re^{-\ri\Omega t}}{\omega-\Omega-\ri\gamma}\Big)\,,}
and from (\ref{a-sol-gen}) we see that 
\FF{\langle \ah\rangle_t\rightarrow\langle \ah\rangle_t^{\rm (\ell c)}=\cF^{\rm (\ell c)}(t)
\label{aav-limit}}
the expectation value for the quantum limit cycle,
which leads to the asymptotic expressions for $\langle x\rangle^{{\rm (\ell c)}}(t)$ and $\langle p\rangle^{{\rm (\ell c)}}(t)$.
 
Alternatively, for real $\Fti(t)$, one can  can start from the equations of motion
\FF{\frac{\rd \langle \xh\rangle}{\rd t}&=&\langle \ph\rangle
-\gamma\,\langle \xh\rangle \label{xav-dot}\\
\frac{\rd \langle \ph\rangle}{\rd t}&=&-\omega^2\,\langle \xh\rangle
-\gamma\,\langle \ph\rangle+ \Fti(t) \label{pav-dot}
}
with $\Fti(t)=\sqrt{2\omega}\fti(t)$. A second time derivative of (\ref{xav-dot}) yields, after inserting (\ref{pav-dot}) and eliminating $\langle \ph\rangle$,
\FF{\frac{\rd^2 \langle \xh\rangle}{\rd t^2}+2\gamma \frac{\rd \langle \xh\rangle}{\rd t} +(\omega^2+\gamma^2)\,\langle \xh\rangle=\Fti(t)\,.\label{dgl-qm}}
We observe, that this differential equation agrees with the classical one in
(\ref{dgl-cl}), where the oscillator frequency $\omega_0$ of 
in the classical equation is replaced by  $\sqrt{\omega^2+\gamma^2}$.
Comparing with section \ref{s-classical} we see that here only the underdamped 
case is realized.\\[2mm]
For a cosine-driving,  the quantum limit cycle can be copied from the classical one in (\ref{x0-cl}) as
\FF{\langle \xh \rangle_t=A_q\cos(\Omega t +\phi_q)\label{x0-qm},}
where the motion follows the external force, shifted by a phase 
\FF{\phi_q=-\arctan\frac{2\gamma\Omega}{\omega^2+\gamma^2-\Omega^2}}
with an amplitude
\FF{A_q=\frac{\Fti_0}{\sqrt{(\omega^2+\gamma^2-\Omega^2)^2+(2\gamma\Omega)^2}}\,.}
In the quantum system, the resonance amplitude is $A_R=\Fti_0/(2\gamma\omega)$ at the
resonance frequency $\Omega_R=\sqrt{\omega^2-\gamma^2}$.
We also note, that in this limit we have
\FF{\langle \ph\rangle^{{\rm (\ell c)}}(t)&=&\rd \langle \xh\rangle/\rd t+\gamma \langle \xh\rangle\nn\\
&=&-\Omega A_q\sin (\Omega t +\phi_q) +\gamma A_q\cos (\Omega t +\phi_q)
\,,\label{p0-qm}}
and in phase space we again obtain an ellipse, namely
\FF{\big(\langle \ph\rangle
-\gamma\,\langle \xh\rangle\big)^2/\Omega^2+\langle \xh\rangle^2=A_q^2\,.\label{limitcycle-qm}}
\vspace*{2mm}

Finally we will evaluate the mean value of $\nh$ by means of 
\FF{&&\frac{\rd \langle \nh\rangle}{\rd t}=\tr \big(\nh\, \frac{\rd \ro}{\rd t}\big)\nn\\[2mm] 
&&=\tr \big(\!-\ri \omega\nh\big[\ad\ah,\ro\big]+\ri \big(\nh\big[\fti^*(t)\ad+\fti(t)\ah,\ro\big]\big) 
\big)\\[2mm] 
&&+\frac{\mu}{2}\,\tr \big(2\nh\ah\ro\ad\!-\!\nh\ad\ah\ro-\nh\ro\ad\ah\big)
\!+\!\frac{\nu}{2}\,\tr \big(2\nh\ad\ro\ah\!-\!\nh\ah\ad\ro\!-\!\nh\ro\ah\ad\big),\nn
}
which can be reduced to 
\FF{\frac{\rd \langle \nh\rangle}{\rd t}=\nu -2\gamma \langle \nh\rangle
+\ri\fti^*(t)\langle \ad\rangle-\ri\fti(t)\langle \ah\rangle\,.\label{nav-t}}
For the force-free oscillator the solution is
\FF{\langle n\rangle_t=\frac{\nu}{2\gamma}
+\Big(n_0-\frac{\nu}{2\gamma}\Big)\,\re^{-2\gamma t}\label{nav-f-0}}
with $\langle n\rangle_t\rightarrow \nu/2\gamma$ for $t\rightarrow \infty$,
which agrees with eq.~(\ref{nav-n-0f-0}) for $n_0=0$.\\[2mm]
For the driven system with $\fti(t)=\fti_0\cos \Omega t$ 
eq.~(\ref{nav-t}) reads 
\FF{\frac{\rd \langle \nh\rangle}{\rd t}=\nu -2\gamma \langle \nh\rangle
+\sqrt{\frac{2}{\omega}}\,f_0\cos \Omega t\,\langle \ph\rangle\,,\label{nav-t-b}}
and, using (\ref{p0-qm}), one can quite easily see that  in the long time limit we have
\FF{\langle \nh\rangle_t^{\rm (\ell c)} &=&\overline{n}+
\frac{\Fti_0A_q}{4\omega}\,\cos(2\Omega t +\phi_q)\label{nav-1}\\
\overline{n}&=&\frac{\nu}{2\gamma}+\frac{\Fti_0A_q}{4\gamma\omega}\,
\Big(\gamma\cos \phi_q-\Omega\sin \phi_q\Big)\,,\label{nav-lc-1}}
which solves the  differential equation (\ref{nav-t}), i.e.~the mean  value of $\nh$
oscillates with twice the forcing frequency, again shifted by
the same phase $\phi_q$, with an average value $\overline{n}$.
\section{A special class of solutions}
\label{s-special}
The fact that the mean values $\langle \nh\rangle_t$,  $\langle \xh\rangle_t$
and  $\langle \ph\rangle_t$ can be described by a closed set
of differential equations, which can be solved without the necessity to
evaluate the full density matrix $\ro (t)$, tremendously simplifies the analysis.
Note that this is no longer true for other operators as for instance  
$\langle \ah^2\rangle_t$.  This simplification is due to the fact that the set
of operators  $\Ih$, $\ah$, $\ad$ and $\ad\ah$ closes under the Lindblad
evolution and the dynamics in this reduced subset simplifies.
We can therefore conjecture that one can
find special solutions generated solely by these
operators expressing the density operator as 
\FF{\ro(t)=\re^{\zz(t)+\vv(t)\ad\ah+\dd(t)\ad+\dd^*(t)\ah}\,,\label{ro-magnus}}
which is Hermitian for  real parameters $\zz$ and $\vv$. This exponential
representation can be rewritten  as an exponential product
\FF{\ro(t)=\re^{\cc(t)}\,\re^{\xb(t)\ad}\re^{\yy(t)\ad\ah}\re^{\xb^*(t)\ah}\,,
\label{ro-exp}}
which is often more convenient. 
The coefficients of both expressions satisfy the disentangling relations 
\FF{\vv=\yy\ ,\quad \dd=\frac{\xb\yy}{\re^\yy-1}\ ,\quad
\zz=\cc+|\xb|^2\,\frac{1+\yy-\re^{\yy}}{(1-\re^\yy)^2}\label{koeff-transf}}
or 
\FF{\yy=\vv\ ,\quad \xb=\frac{\dd}{\vv}\,\Big(\re^\vv-1\Big)\ ,\quad
\cc=\zz-\left|\,\frac{\dd}{\vv}\,\right|^2\Big(1+\vv-\re^{\vv}\Big)\label{koeff-transfx}}
(see, e.g., \cite{Fuji12}, eqs.~(4.4), (4.5)).

Clearly this class of density operators is not general. It covers, however,
some interesting cases. For $\xb=0$ one obtains the canonical distribution,
in the limit $\yy\rightarrow -\infty$ the pure state distribution
$\ro=|\xb\rangle\langle \xb|$ for a coherent state $|\xb\rangle$ and, as
will be shown below, it describes the limit cycle distribution of a 
forced oscillator.
	
In order to verify that such a $\ro(t)$ is actually a solution of the Lindblad equation
we start from the well known operator identities (see, e.g.,~\cite{Loui73})
\FF{\re^{-\xb\ad}\ah\,\re^{\xb\ad}=\ah+\xb\ ,&\quad&  
\re^{-\yy\ad\ah}\ah\,\re^{\yy\ad\ah}=\re^{\yy}\ah\,,\label{op-rel1}\\
\re^{\xb\ah}\ad\,\re^{-\xb\ah}=\ad+\xb\ ,&\quad& 
\re^{\yy\ad\ah}\ad\,\re^{-\yy\ad\ah}=\re^{\yy}\ad\,,\label{op-rel2}}
which imply
\FF{\ah\,\re^{\xb\ad}=\re^{\xb\ad}(\ah+\xb)\ ,&\quad&  
\ah\,\re^{\yy\ad\ah}=\re^{\yy}\,\re^{\yy\ad\ah}\ah\,,\label{reordering1}\\
\re^{\xb\ah}\,\ad=(\ad+\xb)\re^{\xb\ah}\ ,&\quad&
\re^{\yy\ad\ah}\ad=\re^{\yy}\,\ad\,\re^{\yy\ad\ah}\,.\label{reordering2}}
A short calculation using these relations and $\ro$ as given in (\ref{ro-exp}) yields
\FF{\ah\ro=\re^\yy\ro\ah+\xb\ro\ ,\quad \ro\ad 
=\re^\yy\ad\ro+\xb^*\ro\,,\label{a-rho}}
and the time derivative is equal to
\FF{\frac{\rd \ro}{\rd t}&=&\dot c\,\ro+\dot  \xb \ad\ro 
+\dot \yy \re^c\,\re^{\xb\ad}\ad\ah\,\re^{\yy\ad\ah}\re^{\xb^*\ah}
+\dot \xb^*\ro\,\ah\nn\\
&=&\dot c\,\ro+\dot \xb \ad\ro 
+\dot \yy \ad\ro\ah
+\dot \xb^*\ro\,\ah\,,\label{rodot}}
where we have used 
\FF{\re^c\,\re^{\xb\ad}\ad\ah\,\re^{\yy\ad\ah}\re^{\xb^*\ah}
=\re^c\,\ad\,\re^{\xb\ad}\re^{\yy}\,\re^{\yy\ad\ah}\ah\,\re^{\xb^*\ah}
=\re^\yy\ad\ro\,\ah\,.}
As a next step we evaluate ${\mathcal L} \ro$  for the Hamiltonian
(\ref{Hamiltonian}) and reorder the terms starting with the Lindblad
ones using $\ah\ad=\ad\ah+1$:
\FF{&& \frac{\mu}{2}\,\big(2 \ah \ro\ad -\ad \ah\ro- \ro\ad\ah\big)
   +\frac{\nu}{2}\,\big(2 \ad \ro\ah -\ah \ad\ro- \ro\ah\ad\big)\nn\\
   &&\qquad\qquad =\mu \ah\ro\ad+\nu\ad\ro\ah-\gamma'\ad\ah\ro
   -\gamma'\ro\ad\ah-\nu\ro\,.}
We then have
\FF{{\mathcal L} \ro&=&(-\ri\omega +\gamma')\ad\ah\ro
+(\ri\omega-\gamma')\ro\ad\ah-\nu\ro\nn\\
&&\qquad +\ri \fti^*\ad\ro+\ri \fti\ah\ro -\ri\fti^* \ro\ad-\ri \fti\ro\ah
+\mu\ah\ro\ad+\nu\ad\ro\ah\nn\\
&=&(-\ri\omega +\gamma')\ad(\re^\yy\ro\ah+\xb\ro)
+(\ri\omega-\gamma')(\re^\yy\ad\ro+\xb^*\ro)\ah-\nu\ro\nn\\
&&\qquad +\ri \fti^*\ad\ro+\ri \fti(\re^\yy\ro\ah+\xb\ro)
 -\ri\fti^* (\re^\yy\ad\ro+\xb^*\ro)-\ri \fti\ro\ah\nn\\
&&\qquad \qquad +\mu\ah\ro\ad+\nu\ad\ro\ah\,,}
and with relations (\ref{reordering1}) -- (\ref{a-rho}) we find
\FF{\ah\ro\ad&=&\re^\yy\ro\ah\ad+\xb\ro\ad
=\re^\yy\ro(\ad\ah+1)+\xb\re^\yy\ad\ro+|\xb|^2\ro\nn\\
&=&\re^{2\yy}\ad\ro\ah+\re^\yy\xb^*\ro \ah+\xb\re^\yy\ad\ro
+(\re^\yy+|\xb|^2)\ro\,.}
We finally arrive at
\FF{{\mathcal L} \ro &=&
\ri\fti\xb-\ri\fti^*\xb^*-\nu+\mu(\re^\yy+|\xb|^2)\big)\ro\nn\\
&&\quad\quad + \big(\ri\fti^*(1-\re^\yy)+(-\ri\omega-\gamma')\xb 
+\mu \xb\re^\yy\big)\ad\ro\nn\\
&&\quad \quad   +\big(\ri\fti(1-\re^\yy)+(\ri\omega-\gamma')\xb^* 
+\mu\xb^*\re^y\big)\ro\ah\nn\\
&&\quad \quad+\big(\mu\re^{2y} -2\gamma'\re^\yy +\nu\big)\ad\ro\ah\,,\label{Lindblad-2}}
and  in comparison with (\ref{rodot}) we see that the coefficients
satisfy the differential equations
\FF{\dot c &=& \ri\fti\xb-\ri\xb^*-\nu+\mu(\re^\yy+|\xb|^2)\label{coeff-1}\\
\dot \xb &=& \ri\fti^*(1-\re^\yy)+(-\ri\omega-\gamma')\xb 
+\mu \xb\re^\yy\label{coeff-2}\\
\dot \yy&=& -2\gamma' +\mu\re^{\yy}+\nu\re^{-\yy}\,.\label{coeff-3}}
The substitution $u=\re^\yy$ transforms the last equation into
\FF{\dot u=\nu-2\gamma'u+\mu u^2\,,}
which agrees with eq.~(\ref{riccati}). 
A solution
for an initial condition $0<u_0=\uu(0)=\re^{\yy(0)}<1$ is given by
\FF{\uu (t)=\frac{\mu \uu_0-\nu+\nu(1-\uu_0)\re^{2\gamma t}}{\mu \uu_0-\nu+\mu (1-\uu_0)\,\re^{2\gamma t}}\label{sol-ut}}
with $\uu (t)<1$. The function $\uu(t)$ is monotonic, 
decreasing for $u_0>\nu/\mu$ and increasing for $u_0<\nu/\mu$, with $\uu (t)\rightarrow \nu/\mu$
for $t\rightarrow \infty$. For $\nu=0$ we have
\FF{\uu (t)=\frac{\uu_0}{\uu_0+(1-\uu_0)\,\re^{2\gamma t}}\,.\label{sol-ut-nu0}}

Equation (\ref{coeff-2}) for $\xb(t)$ can be simplified
by means of the transformation
$\xb=\xa\,(1-\uu)$, which leads to
\FF{\dot \xa = -(\ri\omega+\gamma)\xa+\ri\fti^*(t)\label{dgl-par-x}}
with initial condition $\xa(0)=\xb(0)/(1-\uu(0))$. 
This agrees with (\ref{a-av-dgl}) for $\langle \ah\rangle_t$ and is solved  by
\FF{\alpha(t)=\re^{-(\ri \omega +\gamma)t}\alpha(0)+\cF(t)\label{alpha-sol-gen}}
with $\cF(t)$ given in (\ref{cFt}).

Integration of (\ref{coeff-1}) yields $c(t)$, i.e.~the normalization factor $Z(t)=\re^{c(t)}$ of
\FF{\ro=Z\,\ro'\quad \textrm{with} \quad \ro'= \re^{\xb\ad}\,\re^{\yy\,\ad\ah}\,\re^{\xb^*\ah}\,,}
but preferably the normalization factor can be obtained in closed form by
evaluating the trace in coherent states $|\xa\rangle$:
\FF{\tr \ro'
&=&\int \frac{\rd^2\xa}{\pi}\,\langle \xa|\re^{\xb\ad}\,\re^{\yy\,\ad\ah}\,\re^{\xb^*\ah} |\xa\rangle \nn\\
&=& \int \frac{\rd^2\xa}{\pi}\,\re^{\xb\xa^*-\bb(t)|\xa|^2+\xb^*\xa}
  =(1/b)\,\re^{\xb^*\xb/b}\label{rosint}}
with 
\FF{\bb(t)=1-\re^{\yy(t)}=1-\uu (t)\,,}
which leads to 
\FF{Z=\bb\,\re^{-{\xb}^*\xb/\bb}\,.\label{Z-norm}}
We see that $\ro$ is normalizable for $\bb>0$, i.e.~$\yy<0$.\\

It should be noted that the normalization factor $Z$ can be employed in the same 
way the partition function in statistical mechanics to generate mean values by parameter differentiation. By means of 
\FF{\frac{\partial \ro\,'}{\partial \xb^*}=\ro\,'\,\ah\ ,\quad \frac{\partial \ro\,'}{\partial \xb}=\ad\,\ro\,'}
one obtains
\FF{ \langle \ah\rangle=-\frac{1}{Z}\,\frac{\partial Z}{\partial \xb^*}
=\frac{\xb}{\bb}\quad \textrm{and} \quad 
\langle \ad\rangle=-\frac{1}{Z}\,\frac{\partial Z}{\partial \xb}
=\frac{\xb^*}{\bb}\label{Z-aav}}
and therefore
\FF{\langle \xh \rangle = \frac{1}{\sqrt{2\omega}}\,\frac{\xb^*+\xb}{\bb}
\quad ,\quad
\langle \ph \rangle =\ri \sqrt{\frac{\omega}{2}}\, \frac{\xb^*-\xb}{\bb}.}
One can also derive expressions for the expectation values
of $\nh=\ad\ah$ or $\ah^{\dagger 2}$ and $\ah^2$ etc. 
using this technique.
As an example, we consider $\nh=\ad\ah$.
Using 
\FF{\frac{\partial \ro'}{\partial \yy}&=& \re^{x\ad}\,\ad\ah\re^{\yy\,\ad\ah}
= \ad\re^{\xb\ad}\,\ah\re^{\yy\,\ad\ah}\nn\\
&=& \ad(\ah-\xb)\re^{\xb\ad}\,\re^{\yy\,\ad\ah}
=\ad\ah\,\ro'-\xb\ad\ah\,\ro'}
one gets
\FF{\langle \nh\rangle&=&-\frac{1}{Z}\,\frac{\partial Z}{\partial \yy} 
+\xb\langle \ad\rangle =\frac{1}{\bb}-1+\frac{|\xb|^2}{\bb^2}\,.\label{Z-nav}}

We have therefore shown that the exponential product (\ref{ro-exp}),
\FF{\ro(t)=Z(t)\,\re^{\xb(t)\ad}\re^{\yy(t)\ad\ah}\re^{\xb^*(t)\ah}
\,,\label{ro-exp-2}}
is form-invariant under Lindblad dynamics. Moreover we have constructed
an explicit solution, which requires only the numerical solution of a single 
differential equation, namely (\ref{dgl-par-x}) for $\xa(t)$:
\FF{\dot \xa = \ri\fti(t)-(\ri\omega+\gamma)\xa\label{dgl-par-x-2}}
with $\xa(0)=\xb(0)/\bb(0)$. This determines the parameter 
\FF{\xb(t)=\xa(t)\bb(t)\,,}
with $b(t)=1-\uu(t)$, where $\uu(t)$ given in  (\ref{sol-ut}) does 
not depend on the force, and the remaining parameter in (\ref{ro-exp-2}) is
given by $\yy(t)=\ln \uu(t)$.
With the normalization  $Z=\bb\,\re^{-|\xa|^2\bb}$ from (\ref{Z-norm})
eq.~(\ref{ro-exp-2}) can be written as
\FF{\ro(t)=\bb(t)\,\re^{-|\xb(t)|^2/b(t)}\,\re^{\xb(t)\ad}\re^{\yy(t)\nh}\re^{\xb^*(t)\ah}
\,\label{ro-exp-3}}
or in pure exponential form by means of the disentangling relation (\ref{koeff-transf})  as
\FF{\ro(t)=\bb(t)\,\re^{\yy(t)\big(|\xa(t)|^2+\nh-\xa(t)\ad-\xa^*(t)\ah\big)}
\,.\label{ro-exp-4}}
The most important expectation values are then 
known analytically, as for example 
\FF{\langle \ah\rangle_t=\xb(t) /\bb(t)=\xa(t)\label{amean-alpha}}
given in eq.~(\ref{Z-aav})). 

From eq.~(\ref{ro-exp-3}) one can also derive a simple equation for the diagonal 
matrix elements of the density operator in coherent states,
\FF{\rho(\xa,t)&=&\langle\xa|\ro(t)|\xa\rangle=
b\,\re^{-|\xb|^2/\bb}\langle \xa|\re^{\xb\ad}\re^{\yy\ad\ah}\re^{\xb^*\ah}|\xa\rangle\nn\\
&=&b\,\re^{-|\xb|^2/\bb}\re^{\xb\ad}\langle \xa|\re^{\yy\ad\ah}|\xa\rangle\re^{\xb^*\xa}\nn\\
&=&b\,\re^{-|\xb|^2/\bb+\xb\xa^*+\bb|\xa|^2+\xb^*\xa}
=b\,\re^{\xb\xa^*+\xb^*\xa},
\label{ro-husimi}}
where we have used
\FF{\langle \xa|\re^{\yy\ad\ah}|\xa\rangle=\re^{-(1-\re^{\yy})\xa^*\xa}\label{exp-loui}}
(see \cite[eq.~(3.3.24)]{Loui73}). 
$\ro(\xa,t)$  is known as the Husimi distribution, a (quasi) density in phase space $(x,p)$
with $x=(\xa^*+\xa)/\sqrt{2\omega}$ and $p=\ri\sqrt{\omega/2\,}\,(\xa^*-\xa)$. 
The Husimi distribution (\ref{ro-husimi}) can be rewritten
in Gaussian form as
\FF{\ro(\xa,t)=\bb(t)
\,\re^{-\bb(t)\,|\xa-\xa(t)\,|^2},\label{ro-husimi-2}}
where $\xa(t)$ describes the motion of the center of the density in phase space, i.e.~the
mean value of $\ah$. The width is given by $1/\sqrt{2\bb(t)}$. 
 For the special case $\nu=0$ we see, using $u(t)$ from (\ref{sol-ut-nu0}),
 that the width parameter is given by
 \FF{\bb (t)=\frac{\bb_0}{\bb_0+(1-\bb_0)\,\re^{-2\gamma t}}\label{sol-bt-nu0}}
 with  $\bb(t)\rightarrow 1$ for $t\rightarrow \infty$.
 
Let us consider some examples in more detail. First,
for a force-free system, eq.~(\ref{dgl-par-x})
is solved by 
\FF{\xa(t)=\xa_0\,\re^{-(\ri \omega+\gamma)t}}
with $\xa_0=\xa(0)$,
and we see that the center of the Gaussian Husimi density distribution (\ref{ro-husimi}) 
traces out the path $\xa(t)$ toward the center at $\xa=0$, whereas the width parameter $\bb(t)$ converges to
$2\gamma/\mu$. This limit agrees, of course, with the steady state distribution (\ref{ro-stst}).
For the harmonically driven system with
$\fti(t)=\fti_0\cos \Omega t$ we can easily verify that $\xa(t)$
is given by 
\FF{\xa(t)=\xa_0\,\re^{-(\ri \omega +\gamma)t}
-\frac{\fti_0}{2}\,\Big(
\frac{\re^{-(\ri \omega +\gamma)t}-\re^{\ri \Omega t}}{\omega+\Omega-\ri\gamma}
+\frac{\re^{-(\ri \omega +\gamma)t}-\re^{-\ri\Omega t}}{\omega-\Omega-\ri\gamma}\Big)
\label{at-sol-2}}
and, according to (\ref{Z-nav}), the expectation value of $\nh$ is equal to
\FF{\langle \nh\rangle_t=\frac{1}{\bb(t)}-1+|\xa(t)|^2\,.\label{nav-2}}
In the long time limit we have
 \FF{\xa (t)\rightarrow \xa^{{\rm (\ell c)}} (t)=
 \frac{\fti_0}{2}\Big(
\frac{\re^{\ri \Omega t}}{\omega+\Omega-\ri\gamma}
+\frac{\re^{-\ri \Omega t}}{\omega-\Omega-\ri\gamma}\Big)}
and therefore, because of $\bb\rightarrow 2\gamma/\mu$,
\FF{\langle \nh\rangle_t=\frac{\nu}{2\gamma}
+\big|\xa^{{\rm (\ell c)}}\big|^2\label{nav-3}}
in agreement with the previous result (\ref{nav-1}).
The density operator converges for long times to the  limit cycle distribution
\FF{\ro^{{\rm (\ell c)}}(t)=\frac{2\gamma}{\mu}\,
\re^{|\alpha^{{\rm (\ell c)}}(t)|^2}
\re^{\beta(t)\ad}\re^{\log(\nu/\mu)\nh}\re^{\beta^*(t)\ah}}
with $\beta(t)=\frac{2\gamma}{\mu}\,\alpha^{{\rm (\ell c)}}(t)$, where we
have used $\bb \rightarrow 2\gamma/\mu$ and $\sigma=\log (1-\bb)
\rightarrow \log (\nu/\mu)$.
Furthermore, the phase space distribution of the limit cycle is simply
\FF{\rho^{{\rm (\ell c)}} (\xa,t)=\frac{2\gamma}{\mu}
\,\re^{-\frac{2\gamma}{\mu}|\,\xa-\xa^{{\rm (\ell c)}} (t) |^2},\label{ro-husimi-3}}
which is a periodic solution with period $T=2\pi/\Omega$, a cyclic steady state
distribution. Note that
this also implies $\langle \ah\rangle^{{\rm (\ell c)}} (t)=\xa^{{\rm (\ell c)}}(t)$ in
agreement  with the result (\ref{aav-limit}). For the special case
$\nu=0$ the limit cycle distribution is a coherent state distribution
\FF{\rho^{{\rm (\ell c)}} (\xa,t)=
\,\re^{-|\,\xa-\xa^{{\rm (\ell c)}} (t) |^2}\,.\label{ro-husimi-4}}

\begin{figure}[h]
\begin{center}
\includegraphics[height=60mm,width=90mm]{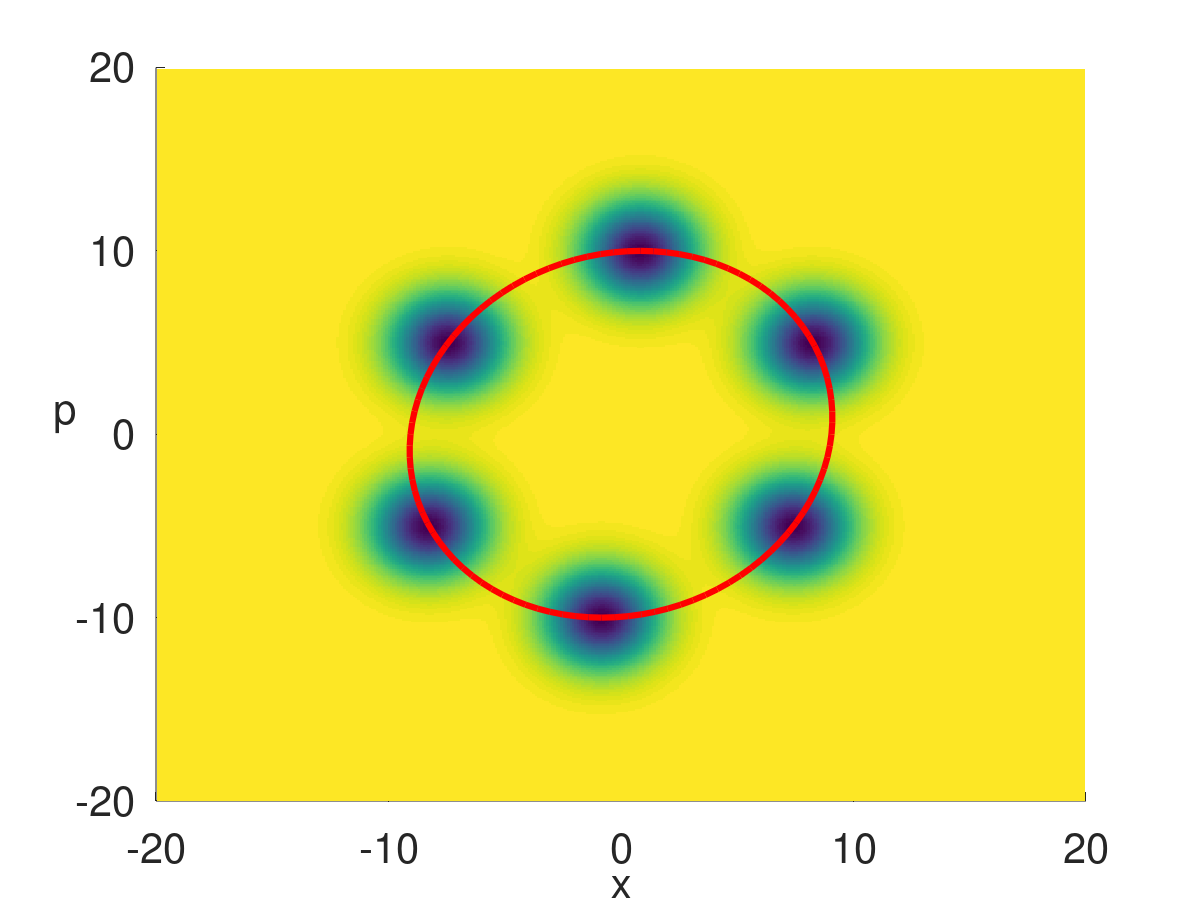}
\end{center}
\caption{\label{fig-husimi} Husimi phase space distribution 
$\ro^{{\rm (\ell c)}}(\alpha,t)$ for the limit cycle given in 
eq.~(\ref{ro-husimi-3}) shown
for  six equidistant times. The center of the distribution follows the ellipse (\ref{limitcycle-qm}) shown as a red curve.}
\end{figure}

Figure \ref{fig-husimi} illustrates the long time dynamics for a system 
with both loss and gain, $\mu=0.6$ and $\nu=0.4$, for $\omega=1.1$,
 and resonant harmonic driving $\Omega=\Omega_R=\sqrt{\omega_0^2-\gamma^2}=1.0954$
with an amplitude $\fti_0=1.4$. The Husimi phase space distribution 
$\ro^{{\rm (\ell c)}}(t)$ for the limit cycle given in eq.~(\ref{ro-husimi-3}) 
is shown in the $(x,p)$-plane \,($\alpha=(\omega x+\ri p)/\sqrt{2\omega}$)\, for six equidistant times in the period $2\pi/\Omega$. 
The center of the distribution follows the ellipse 
(\ref{limitcycle-qm})
where $\langle \xh\rangle^{{\rm (\ell c)}}(t)$ and 
$\langle \ph\rangle^{{\rm (\ell c)}}(t)$ are
given in (\ref{x0-qm}) and (\ref{p0-qm}).

Let us finally remark that 
for $\nu=0$ the
limit cycle distribution turns out to be a pure coherent state, which implies that
any initial distribution is driven into a coherent state. However we know that
a coherent state remains coherent under the evolution, so that its limiting
distribution must be necessarily coherent, with the unavoidable consequence that
the limit state must be coherent.

Closing this section, we briefly look at the entropy
\FF{ S(t)=-\tr \big( \ro(t)\log \ro(t)\big)\,,\label{S-ent}}
which can be most easily evaluated by expressing the density operator in pure
exponential form $\ro=\bb\,\re^{\yy(|\xa|^2+\nh-\xa\ad-\xa^*\ah)}$
as in (\ref{ro-exp-4}). We then have
\FF{ S(t)&=&-\tr \big(\ro\,(\log \bb + \yy (|\xa|^2+\nh-\xa\ad-\xa^*\ah))\big)\nn\\
&=&-\log \bb -\yy \big(|\xa|^2+ \langle \nh\rangle - \xa\langle \ad\rangle - \xa^*\langle \ah\rangle\big)\,.\label{S-ent-2}}
Inserting the expectation values   $\langle \ah\rangle=\xa$,  
$\langle \ad\rangle=\xa^*$ (see eq.~(\ref{amean-alpha})) and 
$\langle \nh\rangle=1/\bb-1+|\xa|^2$  from eqs.~(\ref{Z-aav}) and (\ref{Z-nav}), one
arrives, with  $\bb=1-\re^{\yy}$ and $\yy=\log u$, at
\FF{ S(t)= -\log (1-\uu(t)) -\frac{\uu(t)}{1-\uu(t)}\,\log\uu(t)\,,\label{S-ent-3}}
where $\uu (t)$ is given in closed form in eq.~(\ref{sol-ut}). Because of
$\rd S/\rd \uu>0$ and the monotonicity of $\uu(t)$, the entropy is
monotonically decreasing for $u_0>\nu/\mu$ and increasing for $u_0<\nu/\mu$.
It should be pointed
out, that the parameter $\uu(t)$, and hence the entropy $S(t)$, does 
only depend on the Lindblad parameters $\mu$ and $\nu$ and not 
at all on the driving force. 
In the long time limit $\uu (t)$ approaches $\nu/\mu$
and the resulting value $S_\infty$ is
\FF{S_\infty= -\frac{1}{2\gamma}\Big(\nu\log \nu -\mu\log \mu+
2\gamma\log 2\gamma\Big)\,.\label{S-ent-4}}
In particular we find $S_\infty=0$ for $\nu=0$, in accordance with the fact that
the limit cycle is a coherent state.
Let us also note that the entropy of the (time-dependent) limit cycle distribution
$\ro(t)$  is
constant in time, which is, of course, obvious in view of the fact of the
dynamics of the phase density discussed above. 
More results concerning the concept of entropy in context with the force-free
damped harmonic oscillator can be found in \cite{Isar99}.
\section{Non-Hermitian Hamiltonian dynamics}
\label{s-nh}
In the preceding sections the quantum damped harmonic 
oscillator has been described by a Lindblad master equation. It is, however, also possible to account for
 damping in terms of non-Hermitian Hamiltonians. As an example, the 
 Hamiltonian
\FF{\hat H=\omti \big(\ad\ah+1/2\big)- \fti(t)\,\big(\ad+\ah\big)\label{Hamiltonian-nh}}
with a complex-valued frequency 
\FF{\omti=\omega-\ri\gamma\quad,\quad \omega,\gamma>0\,\label{omti}}
has been considered in \cite{10nhclass} in context with an analysis of a 
semiclassical limit of non-Hermitian quantum dynamics. 
Such a non-Hermitian system can be used to describe a pure loss of particles
and in the following we will compare non-Hermitian and
Lindblad dynamics, in particular for the case of non gain ($\nu=0$).

Let us first consider the simple force-free case $\fti(t)=0$. Then the solution of the
Schr\"odinger equation $\ri \frac{\rd \,}{\rd t}|\psi\rangle=\hat H|\psi\rangle$ is
\FF{|\psi(t)\rangle=\re^{-\ri \hat H t}|\psi_0\rangle
=\re^{-\ri \omti (\ad\ah+1/2) t}|\psi_0\rangle\,.}
The non-Hermitian time evolution does not conserve the norm, which decays as
\FF{\langle\psi(t)|\psi(t)\rangle=\langle\psi_0|\re^{\ri \omti^* (\ad\ah+\frac12) t}
\re^{-\ri \omti (\ad\ah+\frac12) t}|\psi_0\rangle
=\langle\psi_0|\re^{-2\gamma(\ad\ah+\frac12) t}|\psi_0\rangle\,.}
Using (\ref{op-rel1})  we have
\FF{&&\re^{\ri \omti^* (\ad\ah+\frac12) t}\ah\re^{-\ri \omti (\ad\ah+\frac12) t}
=\re^{-\gamma t}\re^{\ri \omti^* \nh t}\ah\re^{-\ri \omti \ad\ah t}\nn\\[1mm]
&&\qquad\quad =\re^{-\gamma t-\ri\omti t}\re^{\ri \omti^* \ad\ah t}\re^{-\ri \omti \ad\ah t}\ah
=\re^{-\gamma t-\ri\omti t}\re^{-2\gamma \ad\ah t}\ah\,,}
and for a coherent initial state $|\xa_0\rangle$ the expectation value of 
$\ah$ is given by
\FF{\langle \ah\rangle_t=\frac{\langle\psi(t)|\ah|\psi(t)\rangle}{\langle\psi(t)|\psi(t)\rangle}
=\xa_0\,\re^{-\ri\omti t}=\xa(t)\,,\label{aav-nh0}}
and the expectation value of $\nh$ is
\FF{\langle \nh\rangle_t=|\xa(t)|^2=n_0\,\re^{-2\gamma t}.\label{nav-nh0}}
with $n_0=|\xa_0|^2$. These results are in perfect agreement with the corresponding
Lindblad expectation values in eq.~(\ref{nav-coh-0}) for $\nu=0$.

For the time-dependent driven system, the analysis is somewhat more involved. A closed 
form solution of the Schr\"odinger equation 
can be found in \cite{10nhclass} and here we will simply copy those results. For a
 coherent initial state $|\xa_0\rangle$, the solution is given by
\FF{|\psi(t)\rangle=\re^{-\ri\omti t/2+A(t)+B(t)\xa_0-|\xa_0|^2/2+|\xa(t)|^2/2}
\,|\xa(t)\rangle,\label{nh-psi}}
where the parameter of the coherent state $|\xa(t)\rangle$ is given by
\FF{\xa(t)=C(t)+\xa_0\,\re^{-\ri\omti t}\label{alphat}}
and  the coefficients $A(t)$, $B(t)$ and $C(t)$  can be found in appendix \ref{ss-nhdyn}.
These terms are, however, only important if one needs to calculate explicitly
the norm $\langle \psi(t)|\psi(t)\rangle$.
If one is only interested in the expectation values 
\FF{\langle \Ah\rangle_t=\frac{\langle\psi(t)|\Ah|\psi(t)\rangle}{\langle\psi(t)|\psi(t)\rangle}}
most of these terms cancel, and we obtain the results
\FF{\langle \ah\rangle_t=
\xa(t)=C(t)+\xa_0\,\re^{-\ri\omti t}\ ,\quad
\langle \nh\rangle_t=|\xa(t)|^2,\label{aav-nav-nh}}
which reduce, of course, for $\fti=0$ to the expressions (\ref{aav-nh0}) and (\ref{nav-nh0}).
These findings are immediately clear if one considers the renormalized density operator
\FF{\ro(t)=\frac{|\psi(t)\rangle\langle \psi(t)|}{\langle\psi(t)|\psi(t)\rangle}
=|\xa(t)\rangle\langle\xa(t)|\,.}
The Husimi phase space distribution can be also calculated easily from the wave function
(\ref{nh-psi}):
\FF{\ro(\xa,t)&=&\frac{|\langle \xa|\psi(t)\rangle|^2}{\langle\psi(t)|\psi(t)\rangle}
=\big|\langle \xa|\xa(t)\rangle\big|^2=\re^{-|\xa-\xa(t)|^2\,.}}
If these results are compared with the Lindblad evolution, we find precise agreement if the
Lindblad parameter $\nu$ is chosen as $\nu=0$ and therefore $\gamma=\mu$. It should
be stressed, however, that this agreement is only valid for an initial coherent 
state, which remains coherent under the Lindblad evolution and, after 
renormalization, also in the non-Hermitian description.
\section{Concluding remarks}
\label{s-con}
In conclusion, we have described the dynamics of an elementary dynamical 
system, a damped and driven harmonic oscillator. This model is almost unavoidable
in teaching physics, however only rarely described in quantum mechanics. Here we
have presented a quite simple analysis using algebraic techniques, which allow
an analytical solution of the Lindblad master equation and a 
calculation of important expectation values in closed form. In particular we have
derived an analytic expression for the asymptotic density operator for harmonic
driving, the limit cycle. Remarkably, for $\nu=0$ this is a coherent  state moving
along the classical limit cycle, a pure state.

The present analysis covers, however, 
only a special class of solutions and the derivation of
a general solution remains a challenging problem for future studies as well as a
deeper study of the interrelation between the Lindblad description and non-Hermitian
Hamiltonians.

\section*{Appendix}
\begin{appendix}
\section{Coherent initial states}
\label{ss-coherent}
Some remarks concerning coherent state distributions may be helpful. As noted
in the beginning of section \ref{s-special}, the exponential product 
\FF{\ro=Z\,\re^{\xb \ad}\re^{\yy \nh}\re^{\xb^{^*}\!\ah}\,,
\label{ro-exp-A2}}
with real $\yy$ and $Z=\bb\,\re^{-|\xb|^2/\bb}$ with $\bb=1-\re^{\yy}$
describes in the limit $\yy\rightarrow -\infty$ the pure state distribution $|\xb\rangle\langle\xb|$ of a 
coherent state $|\xb\rangle$. This is can be easily understood because of
\FF{\lim_{\yy \rightarrow -\infty}\re^{\yy \nh}=\lim_{\yy \rightarrow -\infty}\sum_{n=0}^\infty \re^{\yy n}|n\rangle\langle n|=|0\rangle\langle 0|}
and therefore with $Z\rightarrow \re^{-|\xb|^2/\bb}$ for $\sigma \rightarrow -\infty$
\FF{\lim_{\yy \rightarrow -\infty}\ro=\re^{-|\xb|^2}\re^{\xb \ad}
|0\rangle\langle 0|\re^{\xb^*\ah}=|\xb\rangle\langle\xb|}
(for the last equality see, e.g., \cite[eq.~(3.3.42)]{Loui73}).

An example of such a density operator is found in eq.~(\ref{rho-coh-0})
for the force-free oscillator,  namely 
\FF{\ro(t)=\re^{c(t)}
\,\re^{-\log G(t)\big(\xb(t)\ad+\xb^*(t)\ah -\nh\big)}
\label{rho-coh-A}}
with
\FF{c(t)=|\xa_0|^2\re^{-2\gamma t}\log G(t) +\log(1-G(t))
\ ,\quad \xb(t)=\xa_0\re^{-(\gamma+\ri\omega)t}\,,\label{rho-coh-beta-A}}
where $G(t)$ is given in (\ref{EtGt}), and $\ro(0)=|\xa_0\rangle\langle \xa_0|$.  
By means of the disentangling relations (\ref{koeff-transf}) this can be rewritten 
as an exponential product
\FF{\ro(t)=\re^{c'(t)}\,\re^{\xb(t)(1-G(t))\ad}\,\re^{\log G(t)\,\nh}\,
\re^{\xb^{^*}\!(t)(1-G(t))\ah}}
with $c'(t)=-|\xb(t)|^2(1-G(t))+\log(1-G(t))$. Initially we have $G(0)=0$ and
$\xb(0)=\xa_0$ and from the arguments above we see $\ro(0)=|\xa_0\rangle\langle\xa_0|$.
For $\nu=0$, where we have  $G(t)=0$
for all times, we find in the same way 
\FF{\ro(t)=|\xb(t)\rangle\langle\xb(t)|\label{rho-coh-nu0-A}}
with $\xb(t)=\xa_0\re^{-(\mu/2+\ri\omega)t}$, i.e.~the system stays in a coherent
state for all times.
\section{Details of non-Hermitian dynamics}
\label{ss-nhdyn}
For convenience of the reader we reproduce here some details of the non-Hermitian
dynamics from \cite{10nhclass}. These results  can be derived, e.g., by epressing
the time-evolution operator as an exponential product as in section \ref{s-special} or 
by extending the analysis of the Hermitian forced
harmonic oscillator in \cite[sect.~3.11]{Loui73} to complex frequencies. For a
coherent initial state $|\xa_0\rangle$ this leads to the the wave function  (\ref{nh-psi}),
where  the coefficients $A(t)$, $B(t)$ and $C(t)$  are solutions of the differential
equations
\FF{\ri \dot B=-\fti \re^{-\ri\omti t}\ ,\quad \ri \dot C=\omti C -\fti^*\ ,\quad
\ri \dot A=-\fti C}
with initial conditions $A(0)=B(0)=C(0)=0$, namely
\FF{&&B(t)=-\ri\int_0^t\rd t'\re^{\ri\omti t'}\fti(t') \ ,\quad
C(t)=-\ri\int_0^t\rd t'\re^{\ri\omti (t-t)}\fti^*(t') \nn\\
&&A(t)=\ri\int_0^t\rd t'\,\fti(t') C(t').\label{ABC}}
For harmonic driving force $\fti(t)=\Fti_0\cos \Omega t$ these solutions are given by
\FF{&&B(t)=\frac{-\fti_0}{2(\omti^2-\Omega^2)}\Big\{
(\omti+\Omega)\re^{-\ri(\omti-\Omega)t}+(\omti-\Omega)\re^{-\ri(\omti+\Omega)t}
-2\omti\Big\}\label{Bt}\\
&&C(t)=\frac{\fti_0}{2(\omti^2-\Omega^2)}\Big\{
(\omti-\Omega)\re^{\ri\Omega t}+(\omti+\Omega)\re^{-\ri\Omega t}
-2\omti\re^{-\ri\omti t}\Big\}\label{Ct}\\
&&A(t)=\frac{\fti^2_0}{4(\omti^2-\Omega^2)}\Big\{1+2\ri \omti t
+\frac{\omti+\Omega}{2\Omega}\re^{-\ri(\omti-\Omega)t}-
\frac{\omti-\Omega}{2\Omega}\re^{-\ri(\omti+\Omega)t}\nn\\
&&\qquad \quad +\frac{2\omti}{\omti^2-\Omega^2}\Big((\omti+\Omega)\re^{-\ri(\omti-\Omega)t}+
(\omti-\Omega)\re^{-\ri(\omti+\Omega)t}-2\omti\Big)
\Big\}.\label{At}}
(Note that  the factor $2$ in the denominator in eq.~(\ref{Bt}) is
missing in  \cite{10nhclass}.)
\end{appendix}
%
\bibliographystyle{unsrtot}
\bibliography{abbrev,publko,paper90,paper00,rest}
\end{document}